%Paper: hep-th/9404067
%From: ANNA CERESOLE 3911564 7358 <CERESOLE@polito.it>
%Date: Wed, 13 Apr 1994 15:49:25 GMT+1

%%%%%%%%%%%%%%%%%%%%%%%%%%%%%%%%%%%%%%%%%%%%%%%%%%%%%%%%%%%%%%%%%%%
%                    Real Special Geometry                        %
%                                                                 %
%  M. Bertolini, A. Ceresole, R. D'Auria and S. Ferrara           %
%%%%%%%%%%%%%%%%%%%%%%%%%%%%%%%%%%%%%%%%%%%%%%%%%%%%%%%%%%%%%%%%%%%
\input harvmac
%%%%%%%%%%%%%%%%%%%%%%%%%%%%%%%%%%%%%%%%%%%%%%%%%%%%%%%%%%%%%
\newif\ifdraft

\noblackbox
\catcode`\@=11
\newif\iffrontpage
%%%%%%%%%%%%%%%%%%%%%%%%%%%%%%%%%%%%%%%%%%%%%%%%%%%%%%%%%%%%%%
%%%%% figures
%%%%%%%%%%%%%%%%%%%%%%%%%%%%%%%%%%%%%%%%%%%%%%%%%%%%%%%%%%%%%%
\def\figin{\epsfcheck\figin}\def\figins{\epsfcheck\figins}
\def\epsfcheck{\ifx\epsfbox\UnDeFiNeD
\message{(NO epsf.tex, FIGURES WILL BE IGNORED)}
\gdef\figin##1{\vskip2in}\gdef\figins##1{\hskip.5in}%
\else\message{(FIGURES WILL BE INCLUDED)}%
\gdef\figin##1{##1}\gdef\figins##1{##1}\fi}
\def\DefWarn#1{}
\def\figinsert{\goodbreak\midinsert}
\def\ifig#1#2#3{\DefWarn#1\xdef#1{fig.~\the\figno}
\writedef{#1\leftbracket fig.\noexpand~\the\figno}%
\figinsert\figin{\centerline{#3}}\medskip%
\centerline{\vbox{\baselineskip12pt
\advance\hsize by -1truein\noindent\footnotefont%
\centerline{{\bf Fig.~\the\figno}~#2}}
}\bigskip\endinsert\global\advance\figno by1}
%%%%%%%%%%%%%%%%%%%%%%%%%%%%%%%%%%%%%%%%%%%%%%%%%%%%%%%%%%%%%%
%%%%% sizes, offsets etc
%%%%%%%%%%%%%%%%%%%%%%%%%%%%%%%%%%%%%%%%%%%%%%%%%%%%%%%%%%%%%%
\ifx\answ\bigans
\def\titleft{\titsm}
\magnification=1200\baselineskip=14pt plus 2pt minus 1pt
%
%%%%% unreduced mode: %%%%
%\voffset=0.35truein\hoffset=0.250truein
\advance\hoffset by-0.075truein
\hsize=6.15truein\vsize=600.truept\hsbody=\hsize\hstitle=\hsize
\else\let\lr=L
\def\titleft{\titla}
\magnification=1000\baselineskip=14pt plus 2pt minus 1pt
%
%%%%% reduced mode: %%%%%%%
%\hoffset=-0.5truein\voffset=-.1truein
\hoffset=-.48truein\voffset=-.1truein
\vsize=6.5truein
\hstitle=8.truein\hsbody=4.75truein
\fullhsize=10truein\hsize=\hsbody
\fi
%
%\hstitle=8.truein\hsbody=4.5truein
%\fullhsize=9.75truein\hsize=\hsbody
\parskip=4pt plus 15pt minus 1pt
%%%%%%%%%%%%%%%%%%%%%%%%%%%%%%%%%%%%%%%%%%%%%%%%%%%%%%%%%%%%%%
%%%%%  fonts
%%%%%%%%%%%%%%%%%%%%%%%%%%%%%%%%%%%%%%%%%%%%%%%%%%%%%%%%%%%%%%
%%%%%%%%%%%%%%%%%%%%%%%%%%%%%%%%%%%%%%%%%%%%%%%%%%%%%%%%%%%%%%

\font\titla=cmr10 scaled\magstep3
\font\tenmss=cmss10
\font\absmss=cmss10 scaled\magstep1

\font\twelvebf=cmbx10 scaled\magstep1

\newfam\mssfam
\font\footrm=cmr8  \font\footrms=cmr5
\font\footrmss=cmr5   \font\footi=cmmi8
\font\footis=cmmi5   \font\footiss=cmmi5
\font\footsy=cmsy8   \font\footsys=cmsy5
\font\footsyss=cmsy5   \font\footbf=cmbx8
\font\footmss=cmss8
\def\footfont{\def\rm{\fam0\footrm}
\textfont0=\footrm \scriptfont0=\footrms
\scriptscriptfont0=\footrmss
\textfont1=\footi \scriptfont1=\footis
\scriptscriptfont1=\footiss
\textfont2=\footsy \scriptfont2=\footsys
\scriptscriptfont2=\footsyss
\textfont\itfam=\footi \def\it{\fam\itfam\footi}
\textfont\mssfam=\footmss \def\mss{\fam\mssfam\footmss}
\textfont\bffam=\footbf \def\bf{\fam\bffam\footbf} \rm}
\def\tenpoint{\def\rm{\fam0\tenrm}
\textfont0=\tenrm \scriptfont0=\sevenrm
\scriptscriptfont0=\fiverm
\textfont1=\teni  \scriptfont1=\seveni
\scriptscriptfont1=\fivei
\textfont2=\tensy \scriptfont2=\sevensy
\scriptscriptfont2=\fivesy
\textfont\itfam=\tenit \def\it{\fam\itfam\tenit}
\textfont\mssfam=\tenmss \def\mss{\fam\mssfam\tenmss}
\textfont\bffam=\tenbf \def\bf{\fam\bffam\tenbf} \rm}
\ifx\answ\bigans\def\abstractfont{\tenpoint}\else
\def\abstractfont{\def\rm{\fam0\absrm}
\textfont0=\absrm \scriptfont0=\absrms
\scriptscriptfont0=\absrmss
\textfont1=\absi \scriptfont1=\absis
\scriptscriptfont1=\absiss
\textfont2=\abssy \scriptfont2=\abssys
\scriptscriptfont2=\abssyss
\textfont\itfam=\bigit \def\it{\fam\itfam\bigit}
\textfont\mssfam=\absmss \def\mss{\fam\mssfam\absmss}
\textfont\bffam=\absbf \def\bf{\fam\bffam\absbf}\rm}\fi
%
%%%%%%%%%%%%%%%%%%%%%%%%%%%%%%%%%%%%%%%%%%%%%%%%%%%%%%%%%%%%%%
%%%%% footnotes   (adapted from PHYZZX)
%%%%%%%%%%%%%%%%%%%%%%%%%%%%%%%%%%%%%%%%%%%%%%%%%%%%%%%%%%%%%%
\def\f@@t{\baselineskip10pt\lineskip0pt\lineskiplimit0pt
\bgroup\aftergroup\@foot\let\next}
\setbox\strutbox=\hbox{\vrule height 8.pt depth 3.5pt width\z@}
\def\vfootnote#1{\insert\footins\bgroup
\baselineskip10pt\footfont
\interlinepenalty=\interfootnotelinepenalty
\floatingpenalty=20000
\splittopskip=\ht\strutbox \boxmaxdepth=\dp\strutbox
\leftskip=24pt \rightskip=\z@skip
\parindent=12pt \parfillskip=0pt plus 1fil
\spaceskip=\z@skip \xspaceskip=\z@skip
\Textindent{$#1$}\footstrut\futurelet\next\fo@t}
\def\Textindent#1{\noindent\llap{#1\enspace}\ignorespaces}
\def\footnote#1{\attach{#1}\vfootnote{#1}}%

\def\foot{\attach\footsymbolgen\vfootnote{\footsymbol}}
\let\footsymbol=\star
\newcount\lastf@@t           \lastf@@t=-1
\newcount\footsymbolcount    \footsymbolcount=0
\def\footsymbolgen{\relax\footsym
\global\lastf@@t=\pageno\footsymbol}
\def\footsym{\ifnum\footsymbolcount<0
\global\footsymbolcount=0\fi
{\iffrontpage \else \advance\lastf@@t by 1 \fi
\ifnum\lastf@@t<\pageno \global\footsymbolcount=0
\else \global\advance\footsymbolcount by 1 \fi }
\ifcase\footsymbolcount \fd@f\star\or
\fd@f\dagger\or \fd@f\ast\or
\fd@f\ddagger\or \fd@f\natural\or
\fd@f\diamond\or \fd@f\bullet\or
\fd@f\nabla\else \fd@f\dagger
\global\footsymbolcount=0 \fi }
\def\fd@f#1{\xdef\footsymbol{#1}}
\def\space@ver#1{\let\@sf=\empty \ifmmode #1\else \ifhmode
\edef\@sf{\spacefactor=\the\spacefactor}
\unskip${}#1$\relax\fi\fi}
\def\attach#1{\space@ver{\strut^{\mkern 2mu #1}}\@sf}
%
%%%%%%%%%%%%%%%%%%%%%%%%%%%%%%%%%%%%%%%%%%%%%%%%%%%%%%%%%%%%%%
%%%%% References
%%%%%%%%%%%%%%%%%%%%%%%%%%%%%%%%%%%%%%%%%%%%%%%%%%%%%%%%%%%%%%
\newif\ifnref
\def\rrr#1#2{\relax\ifnref\nref#1{#2}\else\ref#1{#2}\fi}
\def\ldf#1#2{\begingroup\obeylines
\gdef#1{\rrr{#1}{#2}}\endgroup\unskip}
\def\nrf#1{\nreftrue{#1}\nreffalse}
\def\doubref#1#2{\refs{{#1},{#2}}}

\nreffalse
\def\refout{\listrefs}
%
%%%%%%%%%%%%%%%%%%%%%%%%%%%%%%%%%%%%%%%%%%%%%%%%%%%%%%%%%%%%%%
%%%%%%% eq numbering
%%%%%%%%%%%%%%%%%%%%%%%%%%%%%%%%%%%%%%%%%%%%%%%%%%%%%%%%%%%%%%
\def\eqn#1{\xdef #1{(\secsym\the\meqno)}
\writedef{#1\leftbracket#1}%
\global\advance\meqno by1\eqno#1\eqlabeL#1}
\def\eqnalign#1{\xdef #1{(\secsym\the\meqno)}
\writedef{#1\leftbracket#1}%
\global\advance\meqno by1#1\eqlabeL{#1}}
%
%%%%%%%%%%%%%%%%%%%%%%%%%%%%%%%%%%%%%%%%%%%%%%%%%%%%%%%%%%%%%%
%%%%%%  macros for titlepage, marginnotes, etc
%%%%%%%%%%%%%%%%%%%%%%%%%%%%%%%%%%%%%%%%%%%%%%%%%%%%%%%%%%%%%%
\def\chap#1{\global\advance\secno by1\message{(\the\secno\ #1)}
%\ifx\answ\bigans \vfill\eject \else \bigbreak\bigskip \fi  %if
% desired
\global\subsecno=0\eqnres@t\noindent{\twelvebf\the\secno\ #1}
\writetoca{{\secsym} {#1}}\par\nobreak\medskip\nobreak}
%% FOLLOWING LINE CANNOT BE BROKEN BEFORE 70 CHAR
%% FOLLOWING LINE CANNOT BE BROKEN BEFORE 70 CHAR
%% FOLLOWING LINE CANNOT BE BROKEN BEFORE 70 CHAR
\def\eqnres@t{\xdef\secsym{\the\secno.}\global\meqno=1\bigbreak\bigskip}

\def\sequentialequations{\def\eqnres@t{\bigbreak}}\xdef\secsym{}
\global\newcount\subsecno \global\subsecno=0
\def\sect#1{\global\advance\subsecno
by1\message{(\secsym\the\subsecno. #1)}
\ifnum\lastpenalty>9000\else\bigbreak\fi
\noindent{\bf\secsym\the\subsecno\ #1}\writetoca{\string\quad
{\secsym\the\subsecno.} {#1}}\par\nobreak\medskip\nobreak}
%%%%%%%%%%%%%%%%%%%%%%%%%%%%%%%%%%%%%%%%%%%%%%%%%%%%%%%%%%%%%%
%\def\chap#1{\newsec{#1}}
\def\chapter#1{\chap{#1}}
\def\section#1{\sect{#1}}
\def\\{\ifnum\lastpenalty=-10000\relax
\else\hfil\penalty-10000\fi\ignorespaces}
\def\note#1{\leavevmode%
\edef\@@marginsf{\spacefactor=\the\spacefactor\relax}%
\ifdraft\strut\vadjust{%
\hbox to0pt{\hskip\hsize%
\ifx\answ\bigans\hskip.1in\else\hskip .1in\fi%
\vbox to0pt{\vskip-\dp
%\vskip4pt
\strutbox\sevenbf\baselineskip=8pt plus 1pt minus 1pt%
\ifx\answ\bigans\hsize=.7in\else\hsize=.35in\fi%
\tolerance=5000 \hbadness=5000%
\leftskip=0pt \rightskip=0pt \everypar={}%
\raggedright\parskip=0pt \parindent=0pt%
\vskip-\ht\strutbox\noindent\strut#1\par%
\vss}\hss}}\fi\@@marginsf\kern-.01cm}
\def\titlepage{%
\frontpagetrue\nopagenumbers\abstractfont%
\hsize=\hstitle\rightline{\vbox{\baselineskip=10pt%
{\abstractfont\pubnum}}}\pageno=0}
\frontpagefalse
\def\pubnum{}
\def\pdate{\number\month/\number\yearltd}
\def\makefootline{\iffrontpage\vskip .27truein
\line{\the\footline}
%\vskip -.1truein\line{\pdate\hfil}
\vskip -.1truein\leftline{\vbox{\baselineskip=10pt%
{\abstractfont\pdate}}}
\else\vskip.5cm\line{\hss \tenrm $-$ \folio\ $-$ \hss}\fi}
\def\title#1{\vskip .7truecm\titlestyle{\titleft #1}}
\def\titlestyle#1{\par\begingroup \interlinepenalty=9999
\leftskip=0.02\hsize plus 0.23\hsize minus 0.02\hsize
\rightskip=\leftskip \parfillskip=0pt
\hyphenpenalty=9000 \exhyphenpenalty=9000
\tolerance=9999 \pretolerance=9000
\spaceskip=0.333em \xspaceskip=0.5em
\noindent #1\par\endgroup }
\def\autskip{\ifx\answ\bigans\vskip.5truecm\else\vskip.1cm\fi}
\def\author#1{\vskip .7in \centerline{#1}}

\def\address#1{\ifx\answ\bigans\vskip.2truecm
\else\vskip.1cm\fi{\it \centerline{#1}}}
\def\abstract#1{
\vskip .5in\vfil\centerline
{\bf Abstract}\penalty1000
{{\smallskip\ifx\answ\bigans\leftskip 2pc \rightskip 2pc
\else\leftskip 5pc \rightskip 5pc\fi
\noindent\abstractfont \baselineskip=12pt
{#1} \smallskip}}
\penalty-1000}
\def\endpage{\tenpoint\supereject\global\hsize=\hsbody%
\frontpagefalse\footline={\hss\tenrm\folio\hss}}
\def\ack{\goodbreak\vskip2.cm\centerline{{\bf Acknowledgements}}}
%%%%%%%%%%%%%%%%%%%%%%%%%%%%%%%%%%%%%%%%%%%%%%%%%%%%%%%%%%%%%%
\def\a{\alpha}  \def\d{\delta}
\def\e{\epsilon} \def\c{\gamma}

\def\cL{{\cal L}} \def\cM{{\cal M}}

\def\nup#1({Nucl.\ Phys.\ $\us {B#1}$\ (}
\def\plt#1({Phys.\ Lett.\ $\us  {#1}$\ (}
\def\cmp#1({Comm.\ Math.\ Phys.\ $\us  {#1}$\ (}
\def\prp#1({Phys.\ Rep.\ $\us  {#1}$\ (}
\def\prl#1({Phys.\ Rev.\ Lett.\ $\us  {#1}$\ (}
\def\prv#1({Phys.\ Rev.\ $\us  {#1}$\ (}
\def\mpl#1({Mod.\ Phys.\ Let.\ $\us  {A#1}$\ (}
\def\ijmp#1({Int.\ J.\ Mod.\ Phys.\ $\us {A#1}$\ (}
\def\cqg#1({Class.\ Quantum Grav.\ $\us {#1}$\ (}
\def\anp#1({Ann.\ of Phys.\ $\us {#1}$\ (}
\def\tmp#1({Theor.\ Math.\ Phys.\ $\us {#1}$\ (}
\def\rda{R.\ D'Auria}
\def\anna{A.\ Ceresole}
\def\sf{S.\ Ferrara}
\def\tit#1|{{\it #1},\ }
%
%%%%%%%%%%%%%%%%%%%%%%%%%%%%%%%%%%%%%%%%%%%%%%%%%%%%%%%%%%%%%%
%%%%% misc macros %%%%%
%%%%%%%%%%%%%%%%%%%%%%%%%%%%%%%%%%%%%%%%%%%%%%%%%%%%%%%%%%%%%%
\def\ni{\noindent}
\def\tilde{\widetilde}
\def\bar{\overline}
\def\us#1{\underline{#1}}

\def\to{\rightarrow}
\def\notin{\hbox{{$\in$}\kern-.51em\hbox{/}}}

\def\del{\partial}

 \def\ie{{\it i.e.}\ }
\catcode`\@=12

\def\cy{Calabi--Yau\ }
\def\K{K\"ahler\ }
\def\th{{\tilde h}}
%%%%%%%%%%%%%%%%%FRONTPAGE%%%%%%%%%%%%%%%%%%%%%%%%%%%%%%%%%%%%%%%%%%%

\def\aff#1#2{\centerline{$^{#1}${\it #2}}}
\hfill{CERN-TH.7211/94}

\hfill{UCLA/94/TEP/14}

\hfill{POLFIS-TH. 01/94}
\def\pdate{March 1994}
\titlepage
\vskip .5truecm
\title
 {Real Special Geometry\foot{Supported in part by DOE
 grants DE-AC0381-ER50050 and DOE-AT03-88ER40384,Task E.}}
\vskip-.2cm
\author{M.\ Bertolini$^{1}$,
A.\ Ceresole$^{2,3}$, R.\ D'Auria$^{2,3}$ and
S.\ Ferrara$^{4}$}
\vskip2.truecm
\aff1{Dipartimento di Fisica Teorica, Universit\`a di Torino,}
\centerline{\it  Via P. Giuria 1, 10125 Torino, Italy}
\aff2{Dipartimento di Fisica, Politecnico di Torino,}
\centerline{\it  Corso Duca Degli Abruzzi 24, 10129 Torino, Italy}
\aff3{INFN, Sezione di Torino, Italy}
%\line{\hfill}
\aff4{CERN, 1211 Geneva 23, Switzerland}
%\aff5{Department of Physics, UCLA, Los Angeles, USA.}
\line{\hfill}
%%%%%%%%%%%%%%%%%%%%%%%%%%%%%%%%%%%%%%%%%%%%%%%%%%%%%%%%%%%%%%%%%
\vskip-.8 cm
\def\abs
{\ni
We give a coordinate--free description of real manifolds occurring in certain
four--dimensional supergravity theories with antisymmetric tensor fields.
The relevance of the linear multiplets in the compactification of string
and five--brane theories is also discussed.
}
\abstract{\abs}
\vfill
\endpage
\baselineskip=14pt plus 2pt minus 1pt
%%%%%%%%%%%%%%%%%%%%%%%%%%%%%%%%%%%%%%%%%%%%%%%%%%%%%%%%%%%%%%%%%%%%%%%%%
%%%%%%%%%%%%REFERENCES%%%%%%%%%%%%%%%%%%%%%%%%%%%%%%%%%%%%%%%%%%%%%%%%%%
\ldf\laura{L.\ Andrianopoli, unpublished .}
\ldf\rehonomy{for a complete review see L.\ Castellani, \rda\ and P.\ Fr\`e,
\tit Supergravity and Superstrings -- A Geometric Perspective| World
 Scientific (1991) .}
%
%general ref on p--branes:
\ldf\tbrane{
A.\  Strominger, \nup343 (1990) 167;
M.\ J.\ Duff, \cqg5 (1988);
M.\ J.\ Duff and J.\ X.\ Lu, \nup354 (1991) 141;
M.\ J.\ Duff and J.\ X.\ Lu, \prl66 (1991) 1402;
M.\ J.\ Duff and J.\ X.\ Lu, \cqg9 (1992) 1;
M.\ J.\ Duff, R.\ R.\ Khuri and J.\ X.\ Lu, \nup377 (1992) 281;
A.\ Sen, \nup388 (1992) 457 and \plt303B (1993) 22;
A.\ Sen and J.\ H.\ Schwarz, \nup411 (1994) 35;
M.\ J.\ Duff and R.\ R.\ Khuri, \nup411 (1994) 473 .}
\ldf\bin{ P.\ Bin\'{e}truy, \plt315B (1993) 80 .}
\ldf\dfv{
R.\ D'Auria, S.\ Ferrara and M.\ Villasante, \cqg11 (1994) 481 .}
\ldf\siete{ C.\  G.\ Callan, J.\ A.\ Harvey and A.\ Strominger, \nup359
(1991) 611; \nup367 (1991) 60.}
\ldf\rohm{R.\ Rohm and E.\ Witten, \anp170 (1986) 454 ; S.\ J.\ Rey, \prv43
(1990) 526 .}
\ldf\fwz{ S.\ Ferrara, J.\ Wess and B.\ Zumino, \plt51B (1974) 239; W.\ Siegel,
 \plt85B (1979) 333 .}
\ldf\quince{ R.\ Rohm and E.\ Witten, \anp170 (1986) 454;
S.\ J.\ Rey, \prv43 (1990) 526.}
\ldf\trecep{ S.\ Cecotti, S.\ Ferrara and M.\ Villasante, \ijmp2 (1987) 1839 .}
\ldf\lloro{ S.\ Ferrara and M.\  Villasante, \plt186B (1987) 85; P.\
Bin\'{e}truy, G.\ Girardi, R.\ Grimm and M.\ M\"{u}ller, \plt195B (1987) 384 .}
\ldf\sabino{S.\ Ferrara and S.\ Sabharwal, \anp189 (1989) 318 .}
\ldf\Strom  {A.\ Strominger, \cmp133 (1990) 163.}
\ldf\fetal{L.\ Castellani, R.\ D'Auria and S.\ Ferrara, \plt241B (1990) 57;
\cqg1 (1990) 317; R.\ D'Auria, S.\ Ferrara and P.\ Fr\'e, \nup359 (1991) 705.}
\ldf\villa{M.\  Villasante, Phys. Rev. { D45} (1992) 1831;
\anna, \rda, \sf, W.\ Lerche, J.\ Louis and T.\ Regge, \tit Picard--Fuchs
Equations, Special Geometry and Target Space Duality| to appear on \tit
Essays on Mirror Manifolds| vol. II,  S.\ T.\ Yau Editor, International
Press (1993) .}
\ldf\dere{J.\ P.\ Derendinger, F.\ Quevedo and M.\ Quir\'os, \tit The Linear
Multiplet and Quantum Four--dimensional String Effective Actions| preprint
NEIP-93-007, IEM-FT-83/94, hepth 9402007 .}
\ldf\town{M.\ G\"unaydin, G. Sierra and P.\ K.\ Townsend, \nup242 (1984) 244 .}
\ldf\dare{\rda\ and T.\ Regge, \nup195 (1982) 308 .}
%%%%%%%%%%%%%%%%%%%%%%%%%%%%%%%%%%%%%%%%%%%%%%%%%%%%%%%%%%%%%%%%%%%%%%%%%
%%%%%%%%%%%%%%%%%%%%%%%%%%%%%%%BODY%%%%%%%%%%%%%%%%%%%%%%%%%%%%%%%%%%%%%
%%%%%%%%%%%%%%%%%%%%%%%%%%%%%%%%%%%%%%%%%%%%%%%%%%%%%%%%%%%%%%%%%%%%%%%%%
\def\c{\chi}
\def\psi{\Psi}
\def\pb{\bar\psi\ }
\def\nab{{\nabla}}
\def\g{{\gamma}}
\def\cb{\bar\c}
\def\tF{{\tilde F}}
\def\tL{{\tilde L}}

In four--dimensional supergravity theories scalar fields are usually embedded
in linear\fwz\  or chiral multiplets , depending  on whether antisymmetric
tensor fields are present or not. Examples of theories with antisymmetric
tensor fields naturally occur in Kaluza--Klein compactifications of certain
D--dimensional supergravity theories with  $D\geq 6$, or in the low energy
limit of some string and p--brane compactifications. In string theory,
classical examples of linear multiplets $( L,\ B_{\mu\nu},\ \c)$ are the
dilaton multiplet in heterotic strings and the \K class moduli multiplets in
``dual theories'' such as five--brane compactifications
\nrf{\tbrane\bin\dfv}\refs{\tbrane{--}\dfv} .
However, while massless chiral multiplets with an associated continuous
Peccei--Quinn symmetry are classically equivalent to linear multiplets through
a ``duality transformation''
\nrf{\trecep\lloro\dere}\refs{\trecep{--}\dere},
this may be not so at the quantum level, due to the violation of the
Peccei--Quinn symmetry by quantum effects.
The latters comprehend both $\a\prime$ string corrections, \ie quantum effects
on the world-sheet sigma--model, and non perturbative  effects in the string
coupling constant, such as Yang--Mills and gravitational instantons
\nrf{\dare\rohm\siete}\refs{\dare{--}\siete}  .
On the other hand, in supersymmetric compactifications, while $\a\prime$
corrections are expected to be relevant only for configurations for which the
moduli ${\it v.e.v.}$'s are of the order of the string scale, the  breaking of
the space--time axion symmetry may be relevant at a much lower  scale,
especially if we expect that it plays a r\^ole in the supersymmetry breaking
mechanism and in a non--trivial effective dilaton potential which stabilizes
the dilaton field.

Therefore it is of interest, also for physical applications, to treat the
moduli
fields as ``classical'' and the dilaton field as having a non--trivial
superpotential. This is the natural choice in the framework of five--brane
theories\tbrane\ when the \K moduli are associated to linear multiplets
\doubref\bin\dfv\ and the
dilaton (with its pseudoscalar partner) to a chiral multiplet.

 In this letter we show that under such circumstances there is
a new sigma--model geometry associated to the \K class moduli fields $y^i$
which we call ``real special geometry''. Indeed, for particular couplings
of the moduli fields (but not in general), it is related to  special \K
 geometry of \cy compactifications\Strom\fetal
. Like special geometry, in a coordinate free
description real geometry is characterized by a condition on the curvature
tensor
$$
R_{ijkl}=C^m_{\ i[k}\ C_{l]mj}\eqn\curva
$$
where $C_{ijk}$ ($i=1,\ldots, n$) is a completely symmetric tensor.
In a particular set of coordinates $L^I(y)$, it turns out that
$$
\eqalign{C_{IJK} & =\del_I\del_J\del_K\ F(L^I) \cr
G_{IJ} & = \del_I\del_J\ F(L^I)\ ,}\eqn\effe
$$
where $F(L^I)$ is a real function of the scalar fields and $G_{IJ}$ is the
metric tensor. If $C_{IJK}$ is constant, then this geometry is related to the
geometry occurring in five--dimensional supergravity or to the special geometry
of \cy  moduli spaces with
$$
\eqalign{
C_{IJK} & =d_{IJK}\cr
F & =d_{IJK}(T+{\bar T})^I (T+{\bar T})^J (T+{\bar T})^K\ , }\eqn\cala
$$
where $d_{IJK}$ are the intersection numbers of the \cy manifold,
$T^I$ are the  moduli fields of the \K class ($I=1,\ldots,h_{1,1}$)  and $F$
is its ``volume''.

If the $C_{IJK}$ are not constant, the geometry is not
related to special \K geometry, but it is still described by the above
curvature condition.

We now turn to the derivation of real special geometry.
The procedure followed in the sequel is closely analogous to the derivation
of a coordinate free description of special geometry in $D=4$, $N=2$
supergravity\fetal\rehonomy .

 We introduce a set
of $n$ selfinteracting linear multiplets $(L^I(y),\ \c^I,\ B_{\mu\nu}^I )$
, $I=1,\ldots,n$ where $L^I(y)$ are scalar fields functions of the coordinates
$y^i(x^\mu)$ of the sigma--model manifold $\cM_n$ ($x^\mu$ being the four--
dimensional space--time coordinates). $\c^I(x)$ and $B_{\mu\nu}^I(x)$ are
the dilatino and axion fields. We then promote the space--time multiplet to
a superspace multiplet and we introduce the supervielbein basis on superspace
$(V^a,\psi)$ where $V^a,\ a=1,\ldots,4$ is the usual vielbein and $\psi$ is
the gravitino one--form.
For our present purposes it is sufficient to work in global supersymmetry. In
this case, the vielbein $V^a$ and the space--time spin connection satisfy the
zero curvature conditions
$$
\eqalign{
T^a &\equiv D V^a-i \pb \g^a \psi =0\,
\ \ \ DV^a=dV^a-\omega^a_{\ b}\wedge V^b\cr
R^{ab} &= d\omega^{ab}-\omega^a_{\ c}\wedge\omega^c_{\ b}\  .}\eqn\zero
$$

We also introduce on  $\cM_n$ a basis of
``internal'' supervielbein $E^A=E^A_{\ i}\ dy^i$ such that
$$
E^A=E^A_{\ a}\ V^a+\pb \c^A\ ,\eqn\superf
$$
where the supercovariant field strength $E^A_a$ is defined as
$$
E^A_a=(E^A_\mu-\pb_\mu\c^A)V^\mu_{\ a}\, \ \ ,\ \  E^A_\mu=E^A_i{
{\del y^i}\over{\del x^\mu}}\eqn\suuper
$$
and $\c^A$ is related to $\c^I$ by some set of covariant vectors $f^I_{\
A}(y)$:
$$
\eqalign{
\c^I & =f^I_{\ A}\c^A\cr
\c^A & =f^A_{\ I}\c^I\ \ \ \ \ \ \ (f^A_I=(f^I_{\ A})^{-1})\ .}
\eqn\chichi
$$
Together with $E^A$, which may be thought as the curvature of the $y^i$ fields,
we define the curvatures of $\c^A$ and $B^I$ as follows
$$
\eqalign{
\nab \c^A & \equiv d\c^A-\Omega^A_{\ B}\c^B\cr
H^I & \equiv dB^I+ i L^I(y)\pb\g_a\psi V^a\ ,}\eqn\acca
$$
where $\Omega^A_{\ B}$ is the spin connection on $\cM_n$ and
$H^I$ is a three-form on superspace.

The parametrizations of $\nab \c^A$ and $H^I$ are easily found from the
Bianchi identities $\nab E^A=d H^I=0$, and they read
$$
\eqalign{
\nab \c^A = & \nab_a \c^A V^a+{1\over 4}C^A_{\ BC}(\cb^B\c^C+\g_5\cb^B\g_5\c^C)
\psi\cr
&+[{1\over4} h^I_a f^A_I+{i\over2}E^A_{\ a}-{1\over4}
C^A_{BC}\cb^B\g_5\g_a\c^C\g_5]\g^a\psi\cr
H^I= & H^I_{abc}V^a V^b V^c+f^I_{\ A} \cb^A \g_{ab}\psi V^a V^b}\eqn\bianchi
$$
with
$$
f^I_{\ A}=E^i_A\del_i L^I\eqn\effe
$$
and
$$
C^A_{\ BC}=f^A_{\ I}\nab_B f^I_{\ C}\eqn\leci
$$
 symmetric in $BC$.

\ni
We note that the further Bianchi identity
$$
\nab^2 \c^A\equiv R^A_{\ B}\c^B\ ,\eqn\nabdue
$$
where $R^A_{\ B}$ is the curvature two--form, is a true identity since the
linear multiplet gives an off-shell representation of supersymmetry without
auxiliary fields\fwz . For this reason, no constraint on the geometry can be
extracted from \nabdue.
However, geometric constraints can be most naturally obtained by constructing
the superspace lagrangian in the geometric framework\rehonomy, which, in
contrast to  tensor calculus techniques\sabino, gives a lagrangian which is
covariant under reparametrizations of the scalar fields manifold. Applying the
set of rules of \rehonomy, one finds
$$
\eqalign{\cL = &
\ {\it const}\times\left\{ {\tilde E}^A_{\ a}
 (E^A-\cb^A_.\psi_.-\cb^{A.}\psi^.)V^b V^c V^d\e_{abcd}\right .\cr
{}~& -{1\over8}{\tilde E}^A_{\ l} {\tilde E}^A_{\ l} V^a V^b V^c V^d \e_{abcd}
-i (\cb^{A.}\g_a\nab\c^A_.+\cb ^A_.\g_a\nab\c^{A.})V^b V^c V^d \e_{abcd}\cr
{}~& -{3\over2} F_{IJ} \th^I_c [H^J-f^J_{\
A}(\cb^A_.\g_{ab}\psi_.+\cb^{A.}\g_{ab}
\psi^.) V^a V^b] V^c\cr
{}~&
+{1\over{32}} \th^I_c\th^J_cF_{IJ}
 V^a V^b V^c V^d\e_{abcd} +
 3 i E^A(\cb^A_.\g_{ab}\psi_.-\cb^{A.}\g_{ab}\psi^.) V^a V^b\cr
{}~&- 3 i f_{IA} H^I(\cb^A_.\psi_.-\cb^{A.}\psi^.)
 - 6 C_{ABC} f^A_I H^I\cb^{B.}\g_a\c^C_. V^a\cr
{}~&+2 i C_{ABC}(\cb^{C.}\g^d\psi_.+\cb^C_.\g^d\psi^.)
(\cb^A_.\c^B_.+\cb^{A.}\c^{B.}) V^a V^b V^c\e_{abcd}\cr
{}~& \left .+ U_{ABCD}\cb^{A.} \c^{C.}\cb^B_.\c^D_.V^a V^b V^c V^d\e_{abcd}
\right\}\ ,
 }\eqn\lagran
$$
where we have used chiral formalism for a generic spinor field $\lambda$,
$$
\lambda_.={{1+\g_5}\over2}\lambda\ \ \ \
\lambda^.={{1-\g_5}\over2}\lambda\ .
\eqn\chira
$$
\ni
Above, ${\tilde E}^A_{\ a}$ and $\th^I_c$ are auxiliary first order fields
which are identified through their equations of motion with the physical
components (along the vielbein) of $E^A$ and $H^I$\
$$
\eqalign{
{\tilde E}^A_{\ a} &= E^A_{\ a}\cr
\th^I_a &= h^I_a\equiv \e_{abcd}\  H^I_{\ bcd}\ ,
}\eqn\auxilia
$$
\ni
$F_{IJ}$ is a function of the scalar fields $L^I(y(x))$ which appears in the
kinetic term of the axion fields,
 $f_{IA}=\delta_{AB}f^B_{\ I}$ and $U_{ABCD}$ is a four--index tensor so far
undetermined. The superspace equations of motion along the outer
directions (\ie projected on $p$--forms containing at least one $\psi$)
%$$
%\eqalign{
%a_2 & =-i\ ;\ \ a_3=-{3\over2}\ ;\ \ a_5=3 i\ ;\cr
%a_6 & =-3i\ ;\ \ a_7=-6\ ;\ \ a_8=2 i}\eqn\coeffi
%$$
%and also
yield that the scalar field functions satisfy
$$
\eqalign{
\d_{AB} & = F_{IJ} f^I_{\ A} f^J_{\ B}\cr
C_{ABC} & \ \ {\it totally\  symmetric}
}\eqn\scalari
$$
as well as all the numerical coefficients in \lagran .
The form of the tensor $U_{ABCD}$, being a term proportional to the space--time
volume element, cannot be retrieved by an outer projection, but rather it
is determined by a supersymmetry transformation on the lagrangian in a
particular sector. One finds
$$
U_{ABCD}=R_{ABCD}-{3\over2} C^M_{\ BC} C_{ADM}+\nab_D C_{ABC}\eqn\uabcd
$$
\ni
Using these results, we can take the restriction of \lagran\ to space--time
and obtain the four--dimensional supersymmetric lagrangian for the component
fields
$$
\eqalign{
\cL & = {\it const}\  \sqrt{-g}\left\{
 E^A_{\ \mu}E^{A\mu} +2 i (\cb^{A.}
{\not\!\nab}\c^A_.+\cb^A_.{\not\!\nab}\c^{A.})\right .\cr
\  &-{1\over4}F_{IJ}h^I_{\ \mu}h^{J\mu}-2 E^{A\mu}(\cb^A_.\psi_{.\mu}
+\cb^{A.}\psi^._{\mu})\cr
\  &+2
E^{A\mu}(\cb^A_.\g_{\mu\nu}\psi^\nu_.+\cb^{A.}\g_{\mu\nu}\psi^{.\nu})+\cr
\  & i F_{IJ} h^I_\mu f^J_{\ A}\left[\cb^A_.\psi_{.\mu}-\cb^{.A}\psi_{.\mu}-
(\cb^A_.\g_{\mu\nu}\psi_{.\nu}-\cb^{A.}\g_{\mu\nu}\psi^._\nu)\right]\cr
\ &\left . +{i\over2}C_{ABC}\cb^A_.\c^A_.(\cb^{C.}\g_\mu\psi_{.\mu}
+\cb^C_.\g^\mu\psi^._\mu)-{1\over8}U_{ABCD}\cb^{A.}\c^{C.}\cb^B_.\c^D_.
\right\}\ , }\eqn\spaziote
$$
where components along the vielbein of the various (super)covariant
field--strengths have been substituted using the parametrizations
\superf, \bianchi .

Let us exploit the consequences of eqs.\scalari. By differentiating
($\del_C=E^i_C {\del\over{\del y^i}}$) eq.\scalari\ and using \leci\
 we find
$$
\eqalign{
0 &= \del_C F_{IJ} f^I_{\ A}f^J_{\ B}+2 F_{IJ}\nab_C f^I_{\ A} f^J_{\ B}\cr
{}~ &= {{\del F_{IJ}}\over{\del L_K}} f^I_{\ A} f^J_{\ B} f^K_{\ C}
+2 F_{IJ} C^D_{\ CA}f^I_{\ D}f^J_{\ B}\cr
{}~ &= {{\del F_{IJ}}\over{\del L^K}} f^I_{\ A} f^J_{\ B} f^K_{\ C} + 2 C_{ABC}
\ .}
\eqn\differe
$$
\ni
On the other hand, the first of eqs. \scalari\  also implies
$$
{{\del F_{IJ}}\over{\del L^K}}={{\del^2 F_I}\over{\del L^J\del L^K}}=
{{\del^3 F}\over{\del L^I \del L^J \del L^K}}\ ,\eqn\elle
$$
hence
$$
C_{ABC}=-{1\over2}{{\del^3 F}\over{\del L^I\del L^J\del L^K}}f^I_{\ A}
f^J_{\ B}f^K_{\ C}\ .\eqn\lecci
$$
\ni
By covariant differentiation on this equation and using again \leci\
 we also find
$$
\nab_D C_{ABC}=-{1\over2} F_{IJKL} f^I_{\ A}f^J_{\ B}f^K_{\ C}f^L_{\ D}
-{3\over2}F_{IJK}C^L_{\ AD} f^I_{\ L}f^J_{\ B}f^K_{\ C}\eqn\nabci
$$
(with $F_{IJ\ldots}\equiv{\del F\over{\del^I\del^J\ldots}}$), and thus
$$
\nab_{[D}C_{A]BC}=0\ .\eqn\dieci
$$
{}From the definition \leci\ of $C_{ABC}$ one easily deduces
$$
0=\nab_{[D}C_{A]BC}=R_{BCDA}-C^M_{\ B[D}C_{A]CM}\eqn\undici
$$
and finally we obtain the geometric constraint on the curvature
$$
R_{ABCD}=C^M_{\ A[C}C_{D]BM}\ .\eqn\dodici
$$

\ni
Till now we have used the flat vielbein indices $A,B,\ldots$ in the internal
manifold $\cM_n$. If we take  coordinate indices $i,j,\ldots$, then in
special coordinates $y^i=L^I(y)$ we have
$$
f^I_{\ i}=\del_i L^I=\delta^I_{\ i}\eqn\tredici
$$
and thus eqs.\scalari,\lecci\  become
$$
\eqalign{g_{IJ} &= {{\del^2 F}\over{\del L^I \del L^J}}\cr
C_{IJK} &= -{1\over2}{{\del^3 F}\over{\del L^I \del L^J \del L^K}}}\eqn\risul
$$
while the constraint on the curvature becomes
$$
R_{IJKL}=\Gamma^M_{\ I[K}\Gamma_{L]JM}\eqn\levici
$$
{}~$\Gamma^I_{\ JK}$ being the Levi--Civita connection.

It is worth to observe that the above equation is exactly the same found in
ref.
\town\ in the construction of the sigma--model of the scalar fields in $D=5$
supergravity coupled to $N=2$ supermultiplets. There, eq.\levici\ is obtained
as a condition on  the curvature of an $(n-1)$--dimensional hypersurface
$F={\it  const}$ embedded in an $n$--dimensional Riemaniann space $\tilde\cM_n$
and choosing a particular coordinate system. When this theory is dimensionally
reduced down to $D=4$, a new scalar field $\sigma$ appears (from the fifth
component  of the vielbein) and the equation for $F$ becomes $F={\it const}\
\sigma$. By varying $\sigma$, all the space $\tilde\cM_n$ is covered so that it
can be identified with $\cM_n$, while the previously constrained function $F$
becomes actually free and can be identified with our function $F$.

We also remark that we have found the geometric characterization of $\cM_n$
using only the supersymmetric selfinteraction of the linear multiplets without
coupling to supergravity, that is, in a global supersymmetric approach. One
could wonder whether such characterization would change in presence of
supergravity, as it happens in special \K geometry\fetal. There, the curvature
constraint in absence of supergravity
$$
R_{i\bar j k\bar l}=-C_{ikm}C_{\bar j\bar l \bar n} g^{m\bar n} \eqn\speciano
$$
changes to
$$
R_{i\bar j k\bar l}=
g_{i\bar j}g_{k\bar l}+g_{i\bar l} g_{k\bar j}
-C_{ikm}C_{\bar j\bar l \bar n} g^{m\bar n} \eqn\speciasi
$$
when supergravity is turned on. In the above formulae, $C_{ijk}\equiv e^K
W_{ijk}(z)$, where $W_{ijk}$ are the  holomorphic Yukawa couplings and
$K(z,\bar z)$ is the \K potential. However, in the case under investigation,
the constraint \dodici\  is not changed in presence of supergravity. The change
in special \K geometry is due to the fact that the \K manifold of the moduli
fields  in the globally supersymmetric case becomes a \K--Hodge manifold in
presence of supergravity , \ie it acquires the structure of an holomorphic
$U(1)$--bundle with $U(1)$ connection given by $\del_i K(z,\bar z)$. It is
such $U(1)$ gauging which triggers the presence of the extra terms in
\speciasi\  with respect to \speciano . In the present case there is no
superimposed $U(1)$ bundle structure and thus  the constraint \dodici\ remains
unchanged. This has been verified\laura\ by the explicit coupling of
supergravity (in the new minimal framework) to the lagrangian \lagran .

Finally, it is interesting to see what is  the characterization of the special
coordinates $y^i=L^I$  with respect to which the formulas \risul\  hold. It is
easy to show that different sets of special coordinates are related by a
duality transformation, that is by a Legendre transformation with generating
function $F(L)$\doubref\dfv\sabino. Infact, let
$$
\eqalign{
L^I &\to \tL^I\equiv {{\del F}\over{\del L^I}}\cr
\tF(\tL) &= L^I{{\del F}\over{\del L^I}} -F\ .}\eqn\venti
$$
Then $L^I={{\del\tF}\over{\del\tL^I}}$, and therefore
$$
\tF_{IJ}\equiv{{\del^2\tF}\over{\del\tL^I\del\tL^J}}={{\del L^I}
\over{\del\tL^J}}\equiv({{\del\tL^I}\over{\del L^J}})^{-1}=(F^{-1})_{IJ}\ ,
\eqn\ventuno
$$
in agreement with the transformation law of the metric $g_{IJ}\equiv F_{IJ}$
in special coordinates
$$
{\tilde g}_{KL}=g_{IJ}{{\del L^I}\over{\del\tL^K}}{{\del L^J}\over{\del\tL^L}}
\ .\eqn\metrica
$$
Note that the transformation \venti\  also implies
$$
\del_i L^I\to{{\del^2F}\over{\del L^I\del L^J}}\del L^J\ ,\eqn\cincin
$$
that is
$$
f^I_{\ i}\to F_{IJ} f^J_{\ i}\ .\eqn\basta
$$

In this respect, eq. \venti\ is the analogous of $Sp(2n+2)$ transformations
relating different sets of special coordinates in special \K geometry\villa.

A possible generalization of this framework, left to future work,  is to
include chiral and vector multiplets and the coupling to supergravity, which is
of course needed in any study of five--brane compactifications.

\ack It is a pleasure to thank L.\ Andrianopoli and C.\ Vicari for useful
help in the computations during the early stages of this work.

\refout
\end